\documentclass[a4paper,fleqn]{cas-sc}

\usepackage[numbers]{natbib}

\def\tsc#1{\csdef{#1}{\textsc{\lowercase{#1}}\xspace}}
\tsc{WGM}
\tsc{QE}
\tsc{EP}
\tsc{PMS}
\tsc{BEC}
\tsc{DE}

\begin{document}
\let\WriteBookmarks\relax
\def\floatpagepagefraction{1}
\def\textpagefraction{.001}

\shorttitle{Integrating Physics-Based and Data-Driven Approaches for Probabilistic Building Energy Modeling}

\shortauthors{Leandro Von Krannichfeldt et~al.}

\title [mode = title]{Integrating Physics-Based and Data-Driven Approaches for Probabilistic Building Energy Modeling}                      



\author[1,2]{Leandro Von Krannichfeldt}[type=editor,
                        orcid=0000-0001-8563-8086]



\ead{leandro.vonkrannichfeldt@epfl.ch}


\credit{Conceptualization, Methodology, Software, Investigation, Data curation,  Writing - original draft, Writing - review \& editing, Visualization}

\affiliation[1]{organization={EPFL},
    addressline={Route Cantonale}, 
    city={Lausanne},
    postcode={1015}, 
    country={Switzerland}}

\author[3]{Kristina Orehounig}[%
    orcid=0000-0001-6491-7641
   ]

\credit{Supervision, Conceptualization,  Writing - original draft, Writing - review \& editing, Resources}

\affiliation[2]{organization={Swiss Federal Laboratories for Materials Science and Technology (Empa)},
    addressline={Überlandstrasse 126}, 
    city={Dübendorf},
    postcode={8600}, 
    country={Switzerland}}

\affiliation[3]{organization={Vienna University of Technology (TUW)},
    addressline={Karlsplatz 13}, 
    city={Vienna},
    postcode={1040}, 
    country={Austria}}

\author[1]{Olga Fink}[
        orcid=0000-0002-9546-1488
        ]
\ead{olga.fink@epfl.ch}
\cormark[1]

\credit{Supervision, Conceptualization, Writing - original draft, Writing - review \& editing, Resources}

\cortext[cor1]{Corresponding author}

\begin{abstract}
Building energy modeling is a key tool for optimizing the performance of building energy systems. Historically, a wide spectrum of methods has been explored -- ranging from conventional physics-based models to purely data-driven techniques. Recently, hybrid approaches that combine the strengths of both paradigms have gained attention. These include strategies such as learning surrogates for physics-based models, modeling residuals between simulated and observed data, fine-tuning surrogates with real-world measurements, using physics-based outputs as additional inputs for data-driven models, and integrating the physics-based output into the loss function the data-driven model. Despite this progress, two significant research gaps remain. First, most hybrid methods focus on deterministic modeling, often neglecting the inherent uncertainties caused by factors like weather fluctuations and occupant behavior. Second, there has been little systematic comparison within a probabilistic modeling framework. This study addresses these gaps by evaluating five representative hybrid approaches for probabilistic building energy modeling, focusing on quantile predictions of building thermodynamics in a real-world case study. Our results highlight two main findings. First, the performance of hybrid approaches varies across different building room types, but residual learning with a Feedforward Neural Network performs best on average. Notably, the residual approach is the only model that produces physically intuitive predictions when applied to out-of-distribution test data. Second, Quantile Conformal Prediction is an effective procedure for calibrating quantile predictions in case of indoor temperature modeling.
\end{abstract}




\maketitle

\section{Introduction}
\begin{figure}[h]
	\centering
	\includegraphics[width=0.8\linewidth]{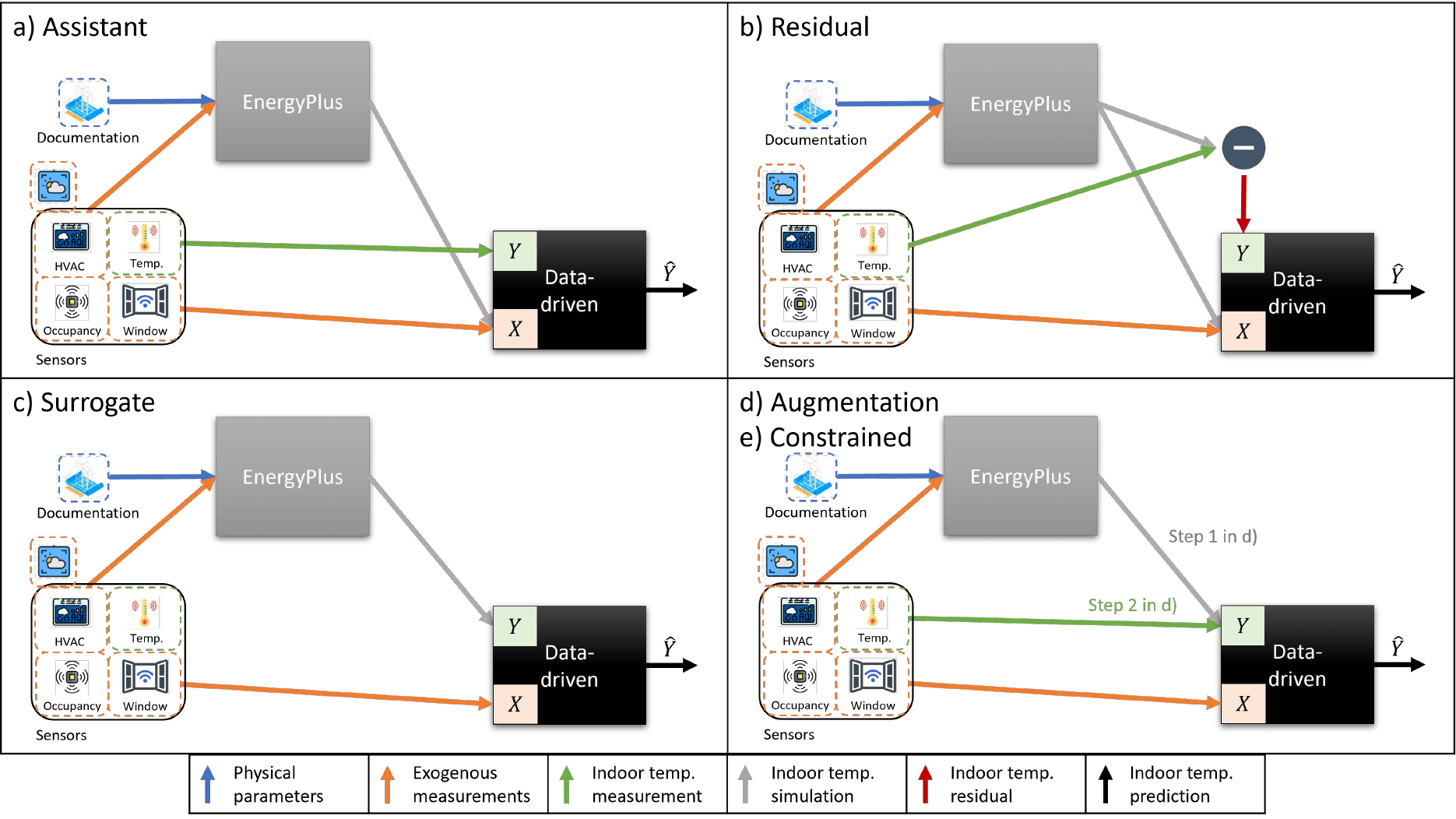}
	\caption{Overview of the five considered hybrid approaches with physics-based EnergyPlus and data-driven sub-models. The two main data sources include building documentation data and building sensor data of various groups. The green exogenous data arrow represents all sensor measurements besides indoor room temperature. While $X$ and $Y$ indicate input features and target variables of the data-driven sub-model, $\hat{Y}$ denotes the final temperature prediction. Note that the Augmentation and Constrained approaches share the same illustration, as both utilize EnergyPlus outputs and temperature data as targets. The key distinction lies in their training procedures: the Constrained approach integrates both targets simultaneously within the loss function, whereas the Augmentation approach first trains on the EnergyPlus outputs and subsequently fine-tunes using real temperature data.}
	\label{fig:hybrid}
\end{figure}

Over the past decade, building operations have accounted for approximately 30\% of global energy use and 26\% of CO\textsubscript{2} emissions \cite{iea_buildings}. More than half of this energy is consumed by systems such as Heating, Ventilation, and Air Conditioning (HVAC), as well as other electrical equipment. This highlights the pressing need for effective energy reduction strategies in building operations. Recent advancements in digital technologies and sensor integration have created new opportunities to apply data analytics and machine learning for improving the efficiency of building energy systems by enabling real-time access. In this context, Building Energy Models (BEMs) are crucial tools for simulating thermodynamic behavior and predicting building energy performance, supporting optimal control decisions in real-world operations. Traditionally, trajectory prediction in BEMs has been framed deterministically, providing a single estimated value per time step \cite{deb_review_2021}. However, these predictions are influenced by multiple sources of uncertainty, including weather variability, occupant behavior, and uncertain physical building parameters such as insulation quality and equipment efficiency \cite{hong_building_2018}. Uncertainties related to the model’s structure or parameters are typically referred to as epistemic, while those arising from external, inherently random factors are referred to as aleatoric. Accounting for both types of uncertainty is essential for robust and reliable decision-making in building energy modeling \cite{yao_state_2021}. Nevertheless, developing models that are both accurate and physically plausible while explicitly capturing uncertainty remains a key challenge. \\
Traditional physics-based BEMs are constructed from first principles, utilizing physical laws and detailed descriptions of building characteristics. These models are typically calibrated using weather data, HVAC operational information, and occupancy schedules \cite{deb_review_2021}. While their development is time- and resource-intensive, they offer  high degrees of physical consistency and interpretability. Popular modeling tools in this domain include the whole building energy simulation tool EnergyPlus (EP) \cite{energyplus} for high-fidelity simulations and thermal Resistance-Capacitance (RC) models \cite{rc_model} for reduced-order modeling. To quantify uncertainty, most studies employ sensitivity analysis, where model outputs are evaluated across input samples generated via Monte Carlo sampling \cite{buechler_probabilistic_2017, guo_uncertainty_2023}. This method addresses epistemic uncertainty by sampling uncertain physical parameters, and captures aleatoric uncertainty by varying weather scenarios. \\
In contrast, data-driven BEMs rely on sensor data to learn statistical or machine learning models that map inputs -—such as operational schedules and environmental conditions—-to outputs like energy consumption or indoor temperature \cite{deb_review_2021}. Approaches for probabilistic modeling range from statistical methods like Bayesian Regression and Quantile Regression to Machine Learning approaches including Ensemble Modeling, Bayesian Neural Networks and Gaussian Processes (GPs) \cite{deb_review_2021, xu_systematic_2025}. While these models often achieve high predictive accuracy, they do not inherently adhere to physical laws. Specifically, uncertainty quantification in data-driven BEMs has been explored using Gaussian Processes \cite{heo_gaussian_2012} or Bayesian LSTM \cite{hannula_bayesian_2025} for building thermodynamics, which can model both epistemic and aleatoric uncertainty. Other approaches include Ensemble Modeling \cite{pachauri_weighted_2023} and Variational Inference \cite{brusaferri_probabilistic_2022} for energy forecasting, Quantile Regression for estimating prediction intervals in energy forecasting \cite{meng_change-point_2020, liu_energy-saving_2024}, and Generative Neural Networks for modeling the full output distribution in building thermodynamics \cite{arpogaus_probabilistic_2025}, mostly focusing on aleatoric uncertainty. \\
In probabilistic prediction with data-driven models, predictive outputs can be miscalibrated for several reasons, such as misspecified priors in Bayesian models or the lack of explicit calibration loss in quantile regression. To improve calibration, model-agnostic frameworks like Conformal Prediction \cite{vovk_algorithmic_2005} have been developed, where predictions are adjusted using a calibration set. Several techniques have been proposed for time series, including Split Conformal Prediction \cite{lei_distribution-free_2018}, Ensemble Batch Prediction Intervals \cite{xu_conformal_2021}, and Adaptive Conformal Inference \cite{gibbs_adaptive_2021}, each differing primarily in how the calibration set is constructed. Split Conformal Prediction, the most widely adopted, uses a dedicated dataset for calibration and has been applied to various machine learning models for deterministic heat load forecasting \cite{fakour_machine_2022, borrotti_quantifying_2024}. Alternatively, Ensemble Batch Prediction Intervals leverage bootstrapped ensembles, particularly in load forecasting with Gradient Boosting Trees \cite{stjelja_building_2024}. However, most existing studies implement Conformal Prediction atop deterministic models, resulting in constant-width prediction intervals and limited conditional coverage. In our work, we use a variant of Split Conformal prediction adapted to quantile regression termed Conformalized Quantile Regression \cite{romano_conformalized_2019}, chosen for its simplicity and suitability for quantile-based modeling.\\
Recently, hybrid approaches that combine physics-based and data-driven methodologies have gained attention, aiming to leverage the strengths of both paradigms \cite{ma_review_2025}. Within this scope, our recent work presents an extensive evaluation of four deterministic hybrid approaches \cite{krannichfeldt_combining_2024}. Building on this foundation, we extend our research in two key directions. First, we introduce probabilistic methods into the hybrid approaches for a comprehensive comparison. Second, we broaden the study with a hybrid approach that incorporates physics-based simulation directly into the data-driven loss function. In this context, we now distinguish five major hybrid approaches in the context of this study: Assistant, Residual, Surrogate, Augmentation and Constrained - all illustrated in Figure \ref{fig:hybrid}. 
The \textbf{Assistant strategy} uses the output of a physics-based model as additional input to a data-driven model, therefore providing more contextual information. This strategy is mainly explored in deterministic prediction context, such as combining EnergyPlus with Linear Regresion \cite{alden_digital_2022} or Gradient Boosting Regression Trees with the IDA-ICE simulation software for energy performance prediction \cite{chen_hybrid-model_2022}.
In the \textbf{Residual strategy}, the data-driven model learns the residuals between actual observed data and output of the physics-based model, thereby capturing unmodelled physical phenomena. In \cite{massa_gray_hybrid_2018} for example, the authors combine a Gaussian Process with an RC-model for thermodynamics modeling, but do not evaluate the posterior predictive distribution despite the bayesian approach. Others integrate the RC-model directly into a Gaussian Process in the spirit of Latent Force Models \cite{ghosh_modeling_2015}.
The \textbf{Surrogate strategy} trains a data-driven model for a low-computation replacement of the physics-based model by using the same input and taking its simulation output as target variable. Examples of this strategy include replacing an RC-model \cite{hossain_evaluating_2019} or an EnergyPlus model \cite{westermann_using_2021} with a Bayesian Neural Network for energy demand prediction. However , these studies do not address the evaluation of the posterior predictive distribution. In the \textbf{Augmentation strategy}, real data is augmented with simulated output from a physics-based model, whereby a data-driven model is trained on the augmented dataset to imbue a physics prior and afterwards adapted to real-world situations. Previous research utilizing this strategy frequently uses EnergyPlus to simulate data for deterministic load prediction with a data-driven model such as LSTM networks \cite{choi_context-aware_2020, ahn_prediction_2022}, among others. In the \textbf{Constrained strategy}, the discrepancy between the physics-based simulation and the final prediction is used as an additional physics-based loss term, thereby regularizing the data-driven model with a physics prior. This approach is predominantly implemented within a deterministic framework, where a Feedforward Neural Network is trained using a regularization loss derived from an RC-model prediction as the physics prior for thermodynamics modeling \cite{gokhale_physics_2022, chen_physics-informed_2023, pavirani_demand_2024}.\\
Despite some advances in hybrid approaches for probabilistic prediction, two notable research gaps remain. First, many hybrid strategies remain largely unexplored within  a probabilistic framework, and even fewer have undergone rigorous  validation  with  real-world data. Second, there is a lack of comprehensive comparisons among different hybrid approaches in probabilistic settings. In this work, we address these gaps by evaluating  five representative hybrid approaches in a real-world case study focused on building thermodynamics -- a domain fundamental to energy-related calculations and characterized by well-separated physical effects.
We adopt a quantile regression methodology for three main reasons. First, we assume uncertainty is primarily aleatoric, stemming from uncontrollable factors such as weather, occupant behavior, and sensor noise. Second, epistemic uncertainty is substantially reduced due to our hybrid approach, which integrates a calibrated high-fidelity BEM simulator, year-round data covering all seasonalities, and multi-source sensor coverage within the building. Third, quantile regression offers flexible temperature distribution modeling without prior assumptions and is computationally efficient, especially when only specific quantiles are needed (e.g., in chance-constrained Model Predictive Control \cite{mohebi_chance_2024}. In addition, we employ Conformalized Quantile Regression \cite{romano_conformalized_2019} to address calibration challenges inherent in temperature quantile regression.
In summary, this paper addresses the identified research gaps through the following three key contributions:

\begin{itemize}
 \item We introduce a probabilistic formulation of five representative hybrid approaches combining physics-based and data-driven models in a quantile regression framework and provide a comprehensive evaluation to enhance the understanding of their specific strengths and weaknesses
  \item We apply Conformalized Quantile Regression to improve the calibration of quantiles as well as predictions intervals and analyze its effect on the hybrid approaches
  \item We validate the five probabilistic hybrid approaches in a real-world case study and analyze their predictions with common metrics for quantiles, prediction intervals as well as full predictive distribution.
\end{itemize}

The remainder of this paper is organized as follows: Section 2 outlines the methodology used in this study. Section 3 introduces the case study set-up and implementation details of the hybrid approaches. Section 4 conducts various experiments and analyzes the results. Finally, Section 5 draws conclusions and provides an
outlook on future research directions.

\section{Methodology}
\subsection{Hybrid probabilistic model}
\begin{figure}[h]
	\centering
	\includegraphics[width=0.7\linewidth]{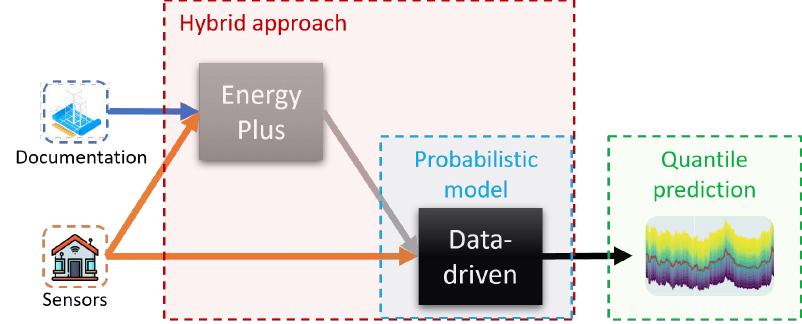}
	\caption{Overview of hybrid framework consisting of physics-based EnergyPlus model and data-driven model.}
	\label{fig:overview_framework}
\end{figure}

In our methodology, we combine physics-based BEM with data-driven techniques within a quantile regression framework for indoor temperature prediction. The BEM computes indoor temperatures based on diverse data inputs, which then serve as features for the data-driven model to estimate specific quantiles of the temperature distribution as illustrated in Figure~\ref{fig:overview_framework}. Depending on the application, these quantile estimates can be used individually or paired to construct prediction intervals. \\
We systematically investigate the five most common hybrid approaches that integrate physics-based and data-driven models for probabilistic building thermodynamics, as illustrated in Figure~\ref{fig:hybrid}. 
The \textbf{Assistant approach} \cite{krannichfeldt_combining_2024} integrates the physics-based temperature simulation as an additional input to the data-driven model, providing complementary information that enhances prediction accuracy. The \textbf{Residual approach} trains a data-driven model to predict the residuals -- i.e., the differences between actual indoor temperature measurements and the outputs of the physics-based simulation. The \textbf{Surrogate approach} involves training a data-driven model to replicate the behavior of the physics-based model, using simulated temperatures as training targets. This enables the surrogate model to replace the original physics-based model for faster inference. The \textbf{Augmentation approach} extends the Surrogate approach by further fine-tuning the data-driven model with real measurement data, thereby improving accuracy and adaptability to real-world conditions. Finally, the \textbf{Constrained approach} integrates a physics prior by adding a loss term that penalizes discrepancies between the physics-based simulation and the model output.
The algebraic formulations for the prediction and loss functions of all five hybrid models are provided below:

\begin{align}
   \mathbf{f}_{assistant}(\mathbf{x}) &= \mathbf{f}_{dd}(\mathbf{x}_{exog}, \mathbf{f}_{EP}(\mathbf{x}_{exog}, \mathbf{x}_{doc})), \quad L_{dd}(\mathbf{y}, \mathbf{f}_{assistant}(\mathbf{x})) \\
    \mathbf{f}_{residual}(\mathbf{x}) &= \mathbf{f}_{dd}(\mathbf{x}, \mathbf{f}_{EP}(\mathbf{x}_{exog}, \mathbf{x}_{doc})), \quad L_{dd}(\mathbf{y} - \mathbf{f}_{EP}(\mathbf{x}_{exog}, \mathbf{x}_{doc}), \mathbf{f}_{residual}(\mathbf{x})) \\
    \mathbf{f}_{surrogate}(\mathbf{x}) &= \mathbf{f}_{dd}(\mathbf{x}_{exog}), \quad L_{dd}(\mathbf{f}_{EP}(\mathbf{x}_{exog}, \mathbf{x}_{doc}), \mathbf{f}_{surrogate}(\mathbf{x})) \\
    \mathbf{f}_{augmentation}(\mathbf{x}) &= \mathbf{f}_{surrogate}(\mathbf{x}_{exog}), \quad L_{dd}(\mathbf{y}, \mathbf{f}_{augmentation}(\mathbf{x})) \\
    \mathbf{f}_{constrained}(\mathbf{x}) &= \mathbf{f}_{dd}(\mathbf{x}_{exog}), \quad L_{dd}(\mathbf{y}, \mathbf{f}_{constrained}(\mathbf{x})) + L_{dd}(\mathbf{f}_{EP}(\mathbf{x}_{exog}, \mathbf{x}_{doc}), \mathbf{f}_{constrained}(\mathbf{x}))
\end{align}

where $\mathbf{f}_{EP}$ and $\mathbf{f}_{dd}$ represent the physics-based EnergyPlus and the data-driven prediction functions, respectively. $\mathbf{x}$ denotes the input features consisting of exogenous measurements $\mathbf{x}_{exog}$ and physical parameters from the building documentation $\mathbf{x}_{doc}$, $\mathbf{y}$ the ground truth measured indoor temperatures, and $L_{dd}$ the loss function used to train the data-driven model.

\subsection{Physics-based model}
\begin{figure}[h]
	\centering
	\includegraphics[width=0.4\linewidth]{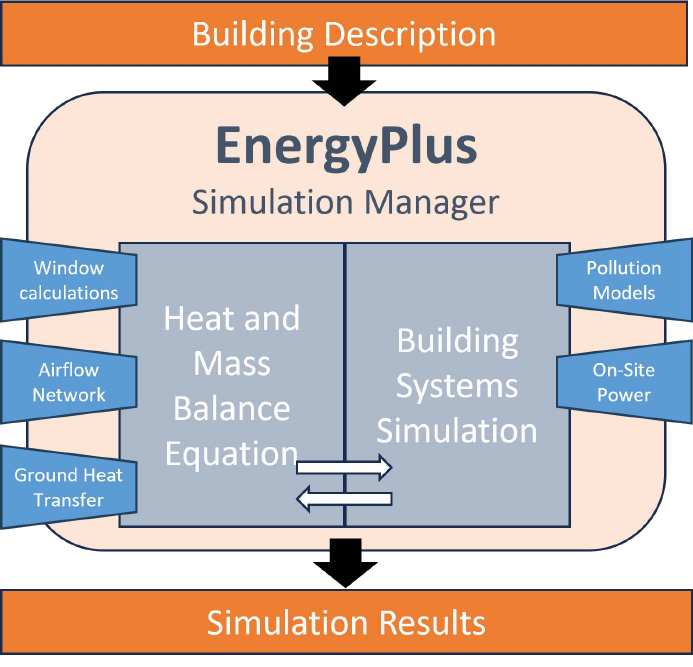}
	\caption{Overview of the EnergyPlus workflow with modules, adapted from \cite{energyplus}}
	\label{fig:overview_energyplus}
\end{figure}

For the physics-based model, we utilize the building simulation software EnergyPlus \cite{energyplus} to develop a comprehensive building energy model.
EnergyPlus simulates the building and its subsystems by solving a system of differential equations distributed across several internal modules as illustrated in Figure~\ref{fig:overview_energyplus}. 
These modules represent essential components, including the surface and air heat balances, cloud conditions and shading conditions, building systems, mass balance, lighting, and window operations. Constructing an EnergyPlus model involves three main steps. First, the building geometry and physical parameters are derived from available documentation and entered into an Intermediate Data Format file, which defines the building's physical and operational characteristics. Second, weather data are prepared in the EnergyPlus weather file format to supply the necessary external climate conditions. Finally, the model is calibrated to ensure accurate and reliable simulations by adjusting physical parameters to minimize a calibration objective, which compares simulation outputs to actual measurements from the building.\\

\subsection{Data-driven quantile models}
For the data-driven quantile models, we evaluate Quantile Regression (QR), Quantile Feedfoward Neural Networks (QNN) and Quantile Random Forest (QRF).
QR is selected as a baseline model for its simplicity and linear structure. QNN is chosen  for its effectiveness in regression tasks and computational efficiency compared to kernel-based methods. QRF, as a tree-based ensemble, is known for its robustness and high accuracy in regression. Both QNN and QRF support multi-output modeling, enabling the simultaneous prediction of thermal dynamics across all rooms with a single unified model.\\
The \textbf{Quantile Regression} \cite{QR} is solved by the following optimization problem:
\begin{equation}
    \mathbf{w}_{q} = \arg\min_{\mathbf{w}} \sum_{i=1}^{N} \rho_{q} (y_i - \mathbf{w}^T \mathbf{x}_i), \quad \rho_{q} (u) =
\begin{cases}
    q u, & u \geq 0, \\
    (q - 1) u, & u < 0.
\end{cases}
\end{equation}

where $N$ is the number of samples, $\mathbf{x}_i$ the $i$-th input, $y_i$ the $i$-th indoor temperature, $\mathbf{w}_q$ the weights for quantile $q$, and $\rho_q$ the pinball loss. \\
The \textbf{Quantile Feedforward Neural Network} predicts quantiles through a multi-layer network:
\begin{align}
    \mathbf{f}(\mathbf{x}) &= \mathbf{z}^{(L)} \circ \mathbf{z}^{(L-1)} \circ ... \circ \mathbf{z}^{(1)}(\mathbf{x}) \\
    \mathbf{z}^{(k)} &= \alpha^{(k)}(\mathbf{W}^{(k)}\mathbf{z}^{(k-1)} + \mathbf{b}^{(k)}) \\
    \mathbf{z}^{(1)} &= \alpha^{(1)}(\mathbf{W}^{(1)}\mathbf{x}+ \mathbf{b}^{(1)})
\end{align}

where, $\mathbf{W}^{(i)}$, $\mathbf{b}^{(i)}$ and $\alpha^{(i)}$ represent the network weights, $i$-th layer's bias term and activation function, respectively. In the equations, the $\circ$-symbol indicates the composition of functions. The network weights are found by minimizing the pinball loss through stochastic approximation of the following optimization problem \cite{bishop_pattern_2006}:

\begin{equation}
    \mathbf{W} = \arg\min_{\mathbf{W}} \dfrac{1}{N} \sum_{i=1}^{N} \rho_q (\mathbf{y}_i - \mathbf{f}(\mathbf{x}_i, \mathbf{W}))
\end{equation}

where $\mathbf{y}_i \in \mathbb{R}^{K}$ is a vector of indoor temperatures from $K$ target rooms. Note that we opt for a multi-output formulation, where $\mathbf{f} (\cdot) \in \mathbb{R}^{Q\cdot K}$ generates an output containing $Q$ quantiles for all target rooms $K$. \\
The \textbf{Quantile Random Forest} \cite{QRF} regression works similar to standard Random Forest by averaging many unbiased tree models to reduce the overall model variance. By denoting a regression tree as $T_b$, the model equation for $B$ number of trees is given as:

\begin{equation}
    \mathbf{f}(\mathbf{x}) = \dfrac{1}{B} \sum_{b=1}^B T_b(\mathbf{x}) \mathbb{1}_{\{y_i \leq y\}}, \quad \quad T_b(\mathbf{x}) = \sum_{j=1}^J \mathbf{c}_j \mathbb{1}_{\{\mathbf{x} \in R_j\}}
\end{equation}

where $R_j$ is a region in the feature space partition, $\mathbf{c}_j$ are the constant values for the region $R_j$ and $\mathbb{1}_{\{\cdot\}}$ represents the indicator function. Moreover, $\mathbf{f}(\cdot)$ represents a distribution function calculated for all $y \in \mathbb{R}$ using the stored target values on each tree leaf. The model parameters are calculated by solving a combinatorial optimization problem:
\begin{equation}
    \mathbf{R}, \mathbf{c} = \arg \min_{\mathbf{R}, \mathbf{c}} \sum_{j=1}^J \sum_{\mathbf{x}_i \in R_j} ||\mathbf{y}_i - \mathbf{c}_j||^2
\end{equation}

\subsection{Conformal Prediction}
Conformal prediction \cite{vovk_algorithmic_2005} is a statistical framework for uncertainty quantification by constructing prediction intervals that offer valid coverage guarantees. The most widely used variant is the split conformal prediction, which relies on three core components. First, a pre-trained prediction model is required. Second, a non-conformity score quantifies the discrepancy between the model’s predictions and the true targets. Third, a calibration data set is used to estimate the conformity scores, leveraging both the ground-truth values and the corresponding model predictions. These scores are then used to adjust the model predictions on new, unseen data. For formal coverage guarantees, it is essential that the calibration and test set are drawn from the same underlying distribution. A variety of non-conformity scores and conformity estimation strategies have been proposed in the literature \cite{lei_distribution-free_2018, xu_conformal_2021}. In our case study, which is focusing on quantile regression, we adopt the Conformalized Quantile Regression framework 
\cite{romano_conformalized_2019} due to its simplicity and its suitability for quantile regression models. The Conformalized Quantile Regression prediction is performed post-training using a pre-trained quantile regression model. This model predicts the lower and upper quantiles for samples in the calibration set, which are then used to compute the non-conformity score $s$, for a given target value $y$ as follows:

\begin{equation}
    s(\mathbf{x}, y) = \max \{f_{\alpha/2}(\mathbf{x})-y, y-f_{1-\alpha/2}(\mathbf{x}) \}
\end{equation}

where $1-\alpha$ denotes the nominal confidence level, with $ \alpha/2$ and $1-\alpha/2$ representing the predicted lower and upper quantiles. The correction factor $\Delta q$ for new predictions is then determined as the $\tfrac{(n+1)(1-\alpha)}{n}$-th empirical quantile of the non-conformity scores:

\begin{equation}
    \Delta q = Quantile(s_1, ..., s_n; \tfrac{(n+1)(1-\alpha)}{n})
\end{equation}

For test predictions, the original quantile outputs are adjusted using $\Delta q$ as follows:

\begin{equation}
    f_{\alpha/2}^{corrected} = f_{\alpha/2} - \Delta q, \qquad f_{1-\alpha/2}^{corrected} = f_{1-\alpha/2} + \Delta q
\end{equation}

\subsection{Evaluation metrics}
We employ three evaluation metrics to assess the performance of our probabilistic regression models. The first is the quantile loss, also known as Pinball Loss (PBL):

\begin{equation}
PBL = \dfrac{1}{N} \sum_{i=1}^{N} \left( \dfrac{1}{Q} \sum_{q} \ell_q (y_i,\hat{y}_q) \right)   \qquad \ell_q(y,\hat{y}_q) = 
\begin{cases}
q (y-\hat{y}_{q}) ,  & y \geq \hat{y}_{q} \\
(q - 1) (y-\hat{y}_{q}) ,   & y < \hat{y}_{q}
\end{cases}
\label{eq:pbl}
\end{equation}

where $q$ is the quantile level, $Q$ the number of quantiles, and $\ell_q$ the quantile loss for each prediction. The summation is performed over all quantiles $q \in \{0.01,\ldots,0.99\}$. \\
When evaluating prediction intervals derived from two quantile forecasts, both reliability and sharpness are important. 
The reliability of the prediction interval is quantified by the Average Coverage Error (ACE) \cite{ACE}, which measures the difference between the empirical coverage and nominal confidence $(1-\alpha)$:

\begin{equation}
ACE = \dfrac{1}{N} \sum_{i=1}^{N} \mathbb{1}_{\{y_i \in [L_i,U_i]\}} - (1-\alpha)
\label{eq:ace}
\end{equation}

The ACE value close to zero indicates that the empirical coverage matches the desired confidence level.\\
Interval sharpness and reliability is measured by the Winkler Score (WKS) \cite{Winkler}:

\begin{equation}
WKS = \dfrac{1}{N} \sum_{i=1}^{N} W(y_i,{L_i,U_i}) \qquad
W(y_i,{L_i,U_i}) = 
\begin{cases}
\delta_i , & L_i \leq y_i \leq U_i \\
\delta_i + 2(L_i - y_i)/\alpha , & y_i < L_i \\
\delta_i + 2(y_i - U_i)/\alpha , & y_i > U_i
\end{cases}	 \qquad	
\end{equation}

where $W$ denotes the Winkler loss, and $\delta_i=U_i-L_i$ is the interval width.

\section{Case study}
\subsection{Dataset}
\begin{figure}[h]
	\centering
	\includegraphics[width=0.5\linewidth]{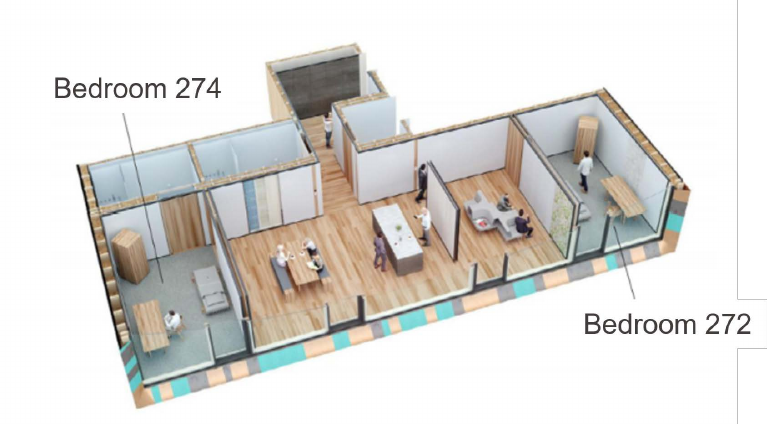}
	\caption{UMAR unit at Empa with two bedrooms, one living room and two bathrooms.}
	\label{fig:umar}
\end{figure}

The aim of the study is to investigate the five hybrid approaches in the context of quantile prediction for indoor temperature in a real-world setting. To this end, the study focuses on the inhabited residential experimental unit Urban Mining and Recycling (UMAR), located at the Swiss Federal Laboratories for Materials Science and Technology (Empa) in Dübendorf, Switzerland \cite{richner_nest_2018}, as shown in Figure~\ref{fig:umar}. Among the various rooms in the unit, we analyze five specific rooms: the living room (R273), two bedrooms (R272 and R274) and two bathrooms (R275 and R276). An overview of the available sensors, which cover weather data, measurements at building level, and room-specific conditions is provided in Table~\ref{table:features}. All sensor quantities are recorded at a 1-minute resolution from January 2020 to December 2021, with less than 1\% missing data. Missing values are addressed through linear interpolation, and the data is then aggregated to a 15-minute resolution. The year 2020 is used as the training set, while 2021 serves as the test set. For conformal prediction, the training set is further split into 80 \% for model training and 20\% for model calibration.

\subsection{Model architecture}
For the \textbf{physics-based models}, we use the widely adopted building simulation software EnergyPlus (EP) \cite{energyplus} . The high-fidelity EP model is a detailed energy simulation of the UMAR unit, constructed from floor plans, construction details, and energy system information, with each room represented as a separate thermal zone. This model is calibrated using sensor data, following the procedure described in \cite{khayatian_benchmarking_2023}. 
It is important to note that both EP models make use of weather features (as listed in ~\ref{table:features}) , but only the high-fidelity EP model incorporates the full set of building and room features.\\
For \textbf{data-driven models} targeting quantile estimation, we evaluate three approaches: Quantile Regression (QR) \cite{QR}, Quantile Feedfoward Neural Network (QNN) and Quantile Random Forest (QRF) \cite{QRF}. QR is chosen for its simplicity, linear structure and strong baseline performance. QNN is included  due to its effectiveness in regression tasks and computational efficiency compared to kernel-based methods. QRF, a tree-based ensemble method, is widely recognized for its robustness and high accuracy in regression settings. Both QNN and QRF are capable of multi-output prediction, enabling simultaneous modeling of all room temperatures within  a unified framework. Input features for QR and QNN are standardized using z-scores. QLR is implemented in CVXPY \cite{cvxpy}, the FFNN is developed using Pytorch \cite{pytorch}, and QRF is based on the Quantile Forest package \cite{quantile-forest}. QLR is formulated as a linear program with an intercept to optimize the quantile loss, as defined in \eqref{eq:pbl}. The QNN comprises two layers, each containing 128 neurons and uses the sigmoid activation function. Training is performed with a batch size of 32, a maximum of 1'000 epochs, early stopping with a patience of 10, and a validation split of 20\% . The QFNN is optimized using the Adam optimizer and the quantile loss function. For QRF, the number of trees is set to 500, the splitting criterion is squared error, the minimum samples per split is 2, and minimum samples per leaf is 1. Unlike standard regular RF, QRF retains all response values at the leaf nodes and computes empirical quantile estimates at inference time. The hyperparameters for both QNN as well as QRF are selected via grid search, where we grid search different two- and three-layer architectures with 64 and 128 neurons for QNN and grid search the number of trees within the range from 100 to 1'000 for QRF.\\
In the case of the \textbf{Augmentation approach}, a fine-tuning approach with early stopping and patience parameter of 3 is used. For the Augmentation-QLR model, the fine-tuning is performed by updating the QLR weights using the Adam optimizer on the real-world dataset. When considering the Augmentation-RF, the fine-tuning is accomplished by utilizing the formerly fitted trees as warm start and adding 100 additional trees for updating on the real-world dataset. The \textbf{Constrained approach} consists of using the physics-based simulation in an additional regularization loss in the loss function of the data-driven model. In our case, we use a regularization constant of 0.1. Considering the Constrained-QLR, we formulated a quantile linear program with regularization loss. For the Constrained-QRF, a physics-constraining regularization is not straightforward to elaborate. To this end, we added the physics-based simulation as additional targets to provide the same information and enable split decisions based on the sum of impurities across all targets.\\
In the probabilistic time series regression framework, the exogenous variables are utilized to predict the indoor temperature quantiles of all rooms at the same time step. Time-lagged data is disregarded as model input to allow direct comparison to EnergyPlus, because it doesn't provide a direct way of selecting historical data integration. For all data-driven and hybrid approaches, we employ quantile estimation of 99 quantiles $q \in \{0.01, ..., 0.99 \}$ and use quantile sorting to ensure monotonically increasing quantile predictions.\\

\begin{table}[h]
    \centering
    \resizebox{\textwidth}{!}{
    \begin{tabular}{@{}ll@{}}
    \toprule
    Feature group    & Feature variables     \\ \midrule
    Datetime    & Season, week (weekday/weekend), daytime (morning/afternoon/evening/night) \\
    Weather     &   Drybulb \& dewpoint temperature, direct \& diffuse solar radiation, rel. humidity, wind direction \& speed\\
    Building    &  Total cooling \& heating mass flows, supply network temperature, air-conditioning mode (cool/heat) \\
    Room        & Mass flow, temperature setpoint, occupancy, window position (closed/open), blinds position (up/down) \\
    \bottomrule
    \end{tabular}}
    \caption{All feature groups with their feature variables}
    \label{table:features}
\end{table}

\section{Results and discussion}

\subsection{Prediction evaluation across rooms}
\begin{figure}[h]
	\centering
	\includegraphics[width=1\linewidth]{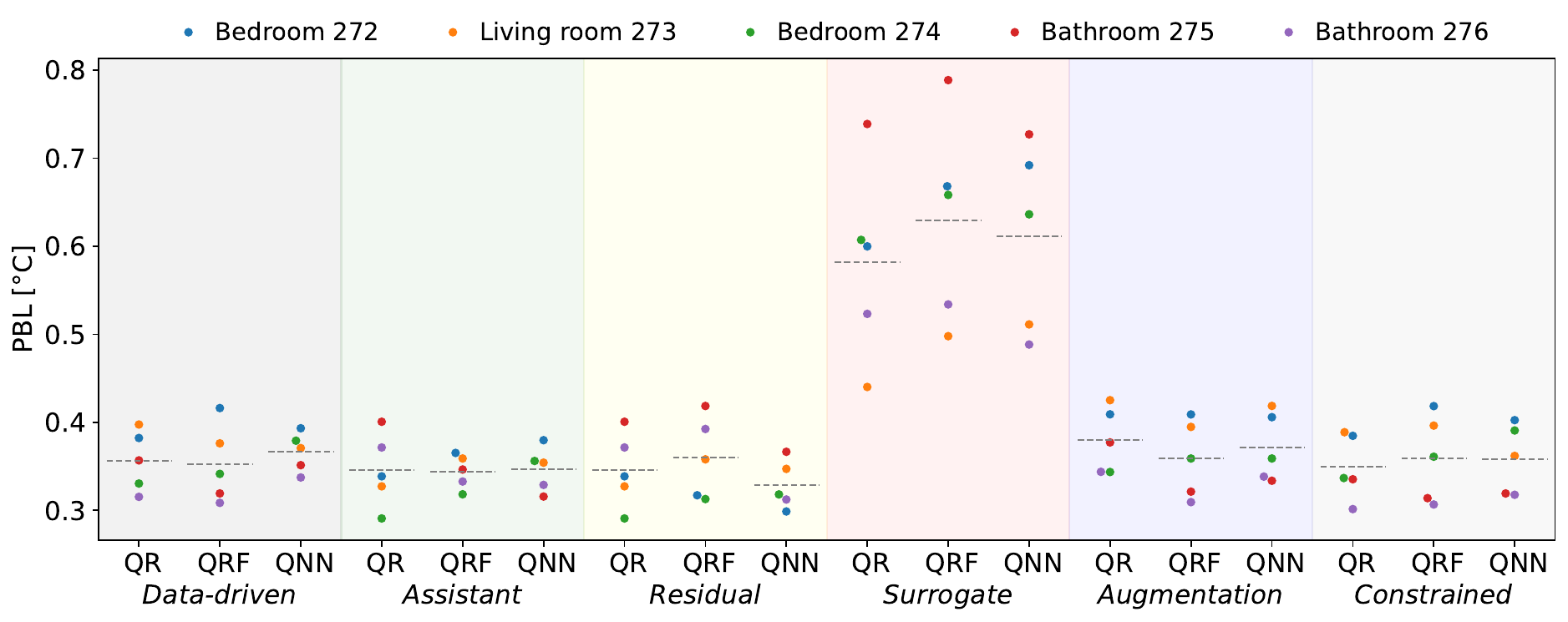}
	\caption{PBL across all 99 quantiles broken down by rooms, for pure data-driven and hybrid approaches. The grey-dotted line indicates the mean PBL across all rooms.}
	\label{fig:room_comparison_PBL}
\end{figure}

In this subsection, we evaluate the performance of various approaches across individual rooms, acknowledging that thermodynamic behavior can differ between room types.
Figure~\ref{fig:room_comparison_PBL} presents the results for the hybrid methods on the PBL metric computed over all 99 quantiles and broken down by room, providing insight into the accuracy of the predicted quantiles of the indoor temperature. The average PBL across rooms is indicated by the dashed grey line. Overall, the Residual approach achieves the lowest average PBL, with Residual-QNN exhibiting the best performance with an average PBL of 0.33. This represents a clear relative improvement of approximately 10\% compared to the purely data-driven QNN, which has an average PBL of 0.37. The Assistant, Augmentation and Constrained approaches demonstrate performance similar to the data-driven baseline. In contrast, the Surrogate approach displays by far the highest PBL among all methods. We also notice that the best performing model differs across rooms. The Residual-QNN shows lowest PBL for bedroom 272, whereas the Residual-QR and Assistant-QR outperform for the bedroom 273 and living room 274. For bathrooms 275 and 275, the Constrained-QRF and Constrained-QR demonstrate the lowest PBL.

\begin{figure}[h]
	\centering
	\includegraphics[width=1\linewidth]{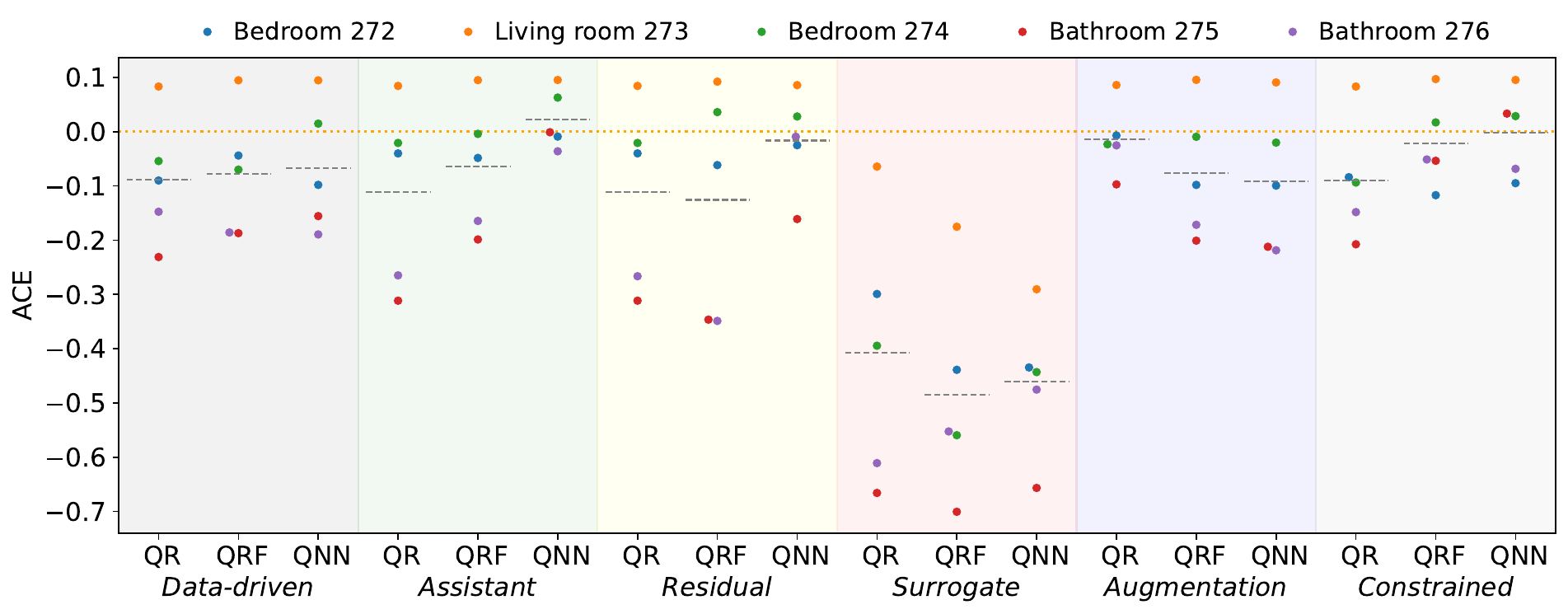}
	\caption{ACE for the 90\% prediction interval broken down by rooms, for pure data-driven and hybrid approaches. The grey-dotted line indicates the mean ACE across all rooms. The orange-dotted line shows the zero axis}
	\label{fig:room_comparison_ACE}
\end{figure}

For uncertainty estimation, it is often useful to consider prediction intervals, defined by upper and lower quantiles that capture a specified range of outcomes. In this study, we focus on the 90\% prediction interval, evaluated using the ACE and WKS metrics. Figure~\ref{fig:room_comparison_ACE}, shows the ACE for the 90\% interval, which is defined by the 5\% and 95\% quantile predictions. ACE quantifies the difference between the nominal confidence level (90\%) and the empirical coverage achieved by the model. Among all models, the Constrained-QNN demonstrates the highest calibration accuracy, with Augmented-QR and Residual-QNN also performing well.
Notably, the hybrid QNN approaches (Assistant, Residual and Constrained) achieve higher reliability than the purely data-driven models, highlighting the benefit of incorporating high-fidelity physics-based simulation.
While bedroom 274 appears to be the easiest target, yielding the lowest ACE, most models struggle with bathrooms 275 and 276. This reflects the greater challenge in modeling distinct thermodynamics of bathrooms, likely due to their ventilation systems and intermittent, intensive usage. 

\begin{figure}[h]
	\centering
	\includegraphics[width=1\linewidth]{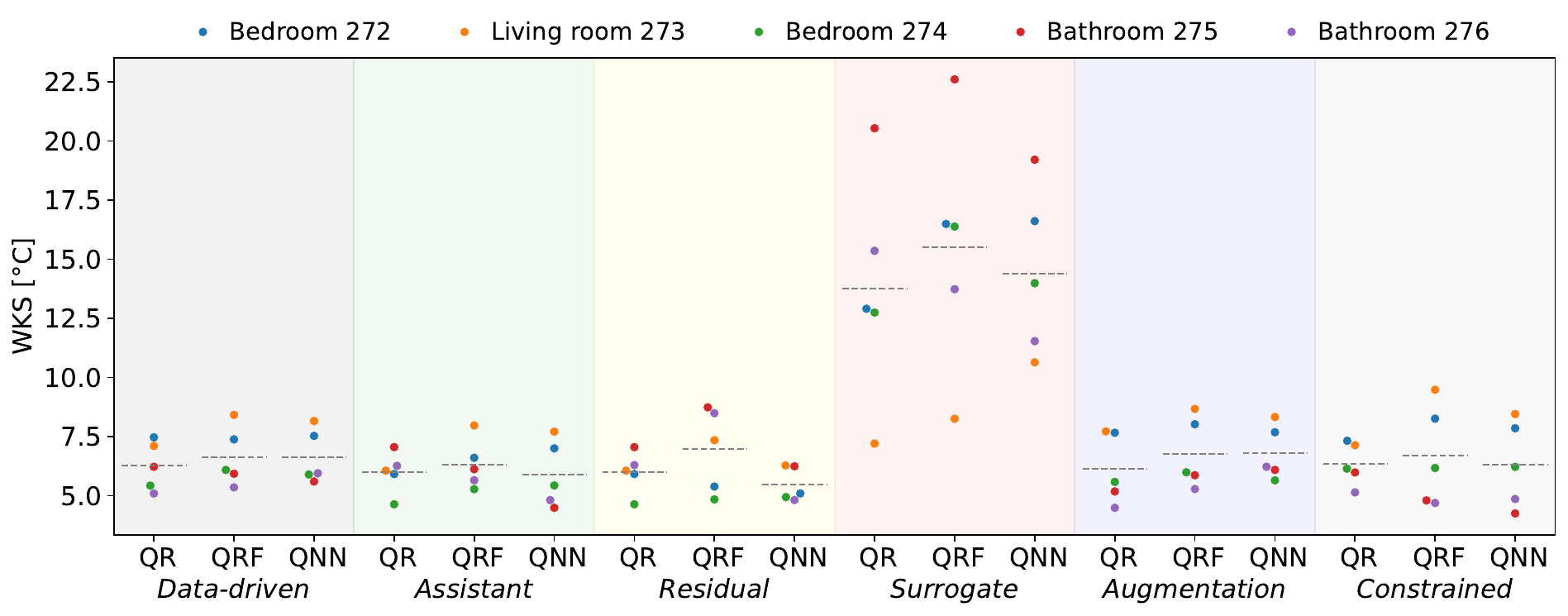}
	\caption{WKS for the 90\% prediction interval broken down by rooms, for pure data-driven and hybrid approaches. The grey-dotted line indicates the mean WKS across all rooms.}
	\label{fig:room_comparison_WKS}
\end{figure}

Even though several hybrid approaches exhibit a low ACE, a model could just widen the prediction interval to reduce ACE while decreasing informativeness. To this end, we evaluate prediction interval sharpness and reliability combined with the WKS in Figure~\ref{fig:room_comparison_WKS} for the 90 \% prediction interval. We observe that the Constrained-QNN does not display the lowest mean WKS, despite having the lowest ACE. This result indicates that, on average, the prediction intervals have become wider. The best-performing model in terms of WKS is the Residual-QNN, showing that this hybrid approach achieves the best balance between sharpness and reliability. In contrast, the surrogate approach consistently performs the worst, while the data-driven methods show performance comparable to the remaining hybrid approaches.

\begin{figure}[h]
	\centering
	\includegraphics[width=1\linewidth]{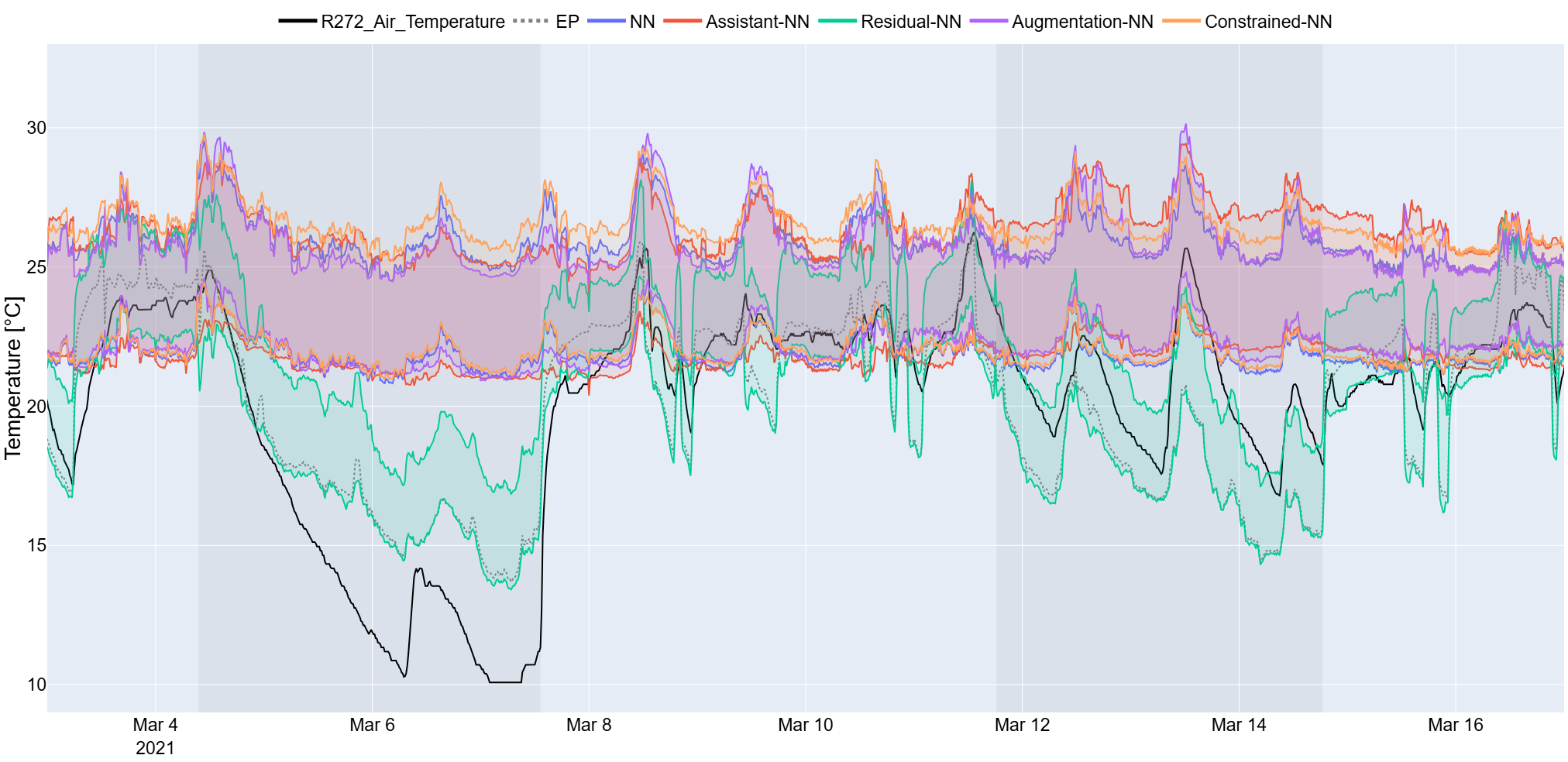}
	\caption{Visualization of the 5\%- and 95\%-quantile enclosing a 90\% prediction interval for a selection of hybrid approaches using the QNN sub-model in the case of Room 272. The forecast of the Surrogate approach is omitted for better visibility. The grey shaded areas indicate the extended period of window openings.}
	\label{fig:forecast_window_mqfnn}
\end{figure}

One reason for the better performance of the Residual approach is its robustness to out-of-distribution scenarios. This is illustrated in Figure~\ref{fig:forecast_window_mqfnn}, which displays the 90\% prediction intervals for the hybrid QRF approaches in room 272. The grey shaded area regions mark periods of prolonged window openings in the test set around March 6th and March 14th, events that are not represented to such an extent in the training data. Notably, while all models have access to EnergyPlus predictions during training, only the Residual approach effectively leverages the physics-based predictions to capture temperature drops during these extended window openings. In contrast, the purely data-driven and other hybrid approaches fail to detect these deviations. Furthermore, the Residual approach consistently produces sharper prediction intervals than the other methods. This may be attributed to the fact that learning the Residual focuses on modeling a smaller range of variation relative to the full thermodynamic behavior, resulting in tighter and more precise quantile predictions.

\subsection{Physics-constrained loss sensitivity}

\begin{figure}[h]
	\centering
	\includegraphics[width=0.7\linewidth]{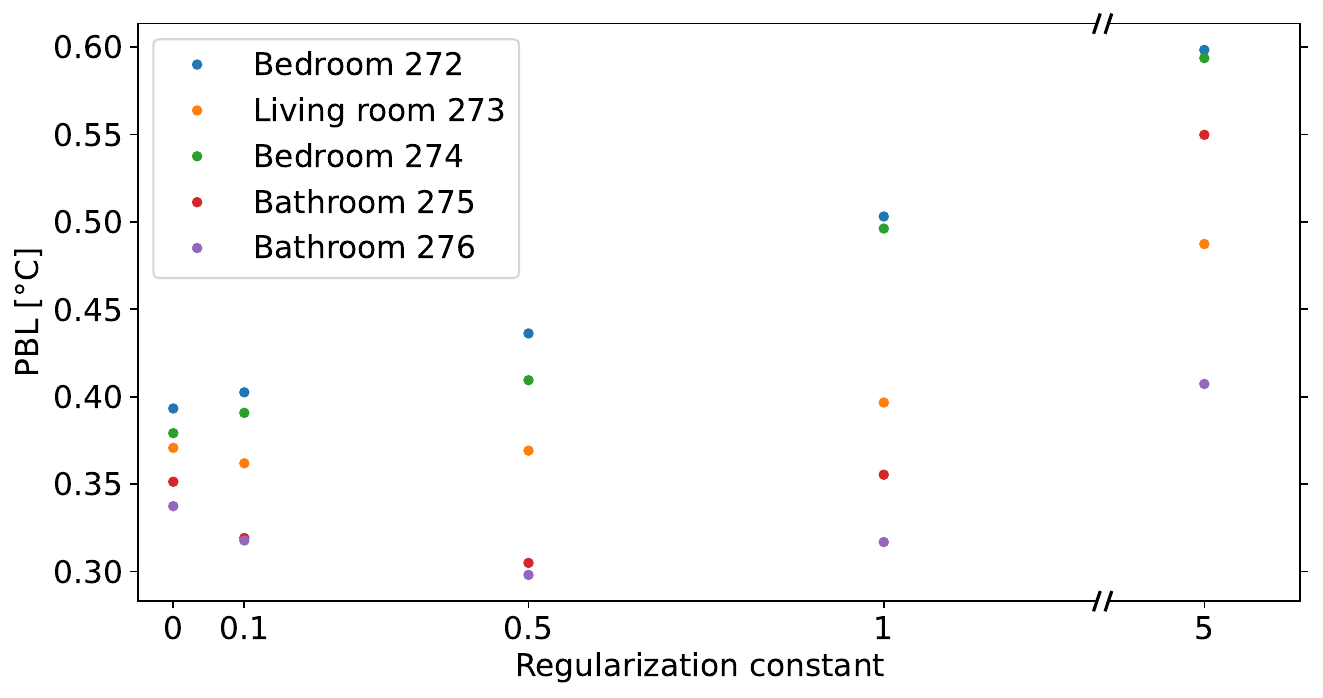}
	\caption{Sensitivity analysis of the physics regularization constant of the QNN physics-constrained hybrid approach. While 0 indicates a total loss function fully consisting of the data-driven loss, the regularization constants 0.1, 0.5, 1 and 5 represent a total loss with 10\%, 33.33\%, 50\% and 66.67\% physics-based loss share.}
	\label{fig:pinn_sensitivity}
\end{figure}

In this subsection, we examine the Constrained approach in greater detail for two main reasons. First, it is the only hybrid approach that incorporates an explicit trade-off parameter to balance the contributions of the physics-based and data-driven components. Second, we aim to investigate whether the proportion of incorporated physics-based information influences the resulting uncertainty estimates.
To investigate this, we conduct a sensitivity analysis of the regularization constant governing the physics-based loss in the Constrained approach, as shown in Figure~\ref{fig:pinn_sensitivity}. Two main trends emerge from the results.
First, increasing the constraint constant beyond 0.5 leads to a rise in PBL. This suggests that as the physics-based loss becomes the dominant driver during training, performance deteriorates. This is expected since the physics-based model, only calibrated on training data, remains an approximation of the true temperature. When the constraint constant exceeds 1, meaning the physics-based loss contributes more than 50\% to the total loss, the predictive performance further degrades.
Second, including the physics-based loss does not yield noticeable benefits for bedrooms. However, there is a clear improvement for bathrooms, especially when the constant is set to 0.5. This may be because bathroom dynamics are more complex and difficult to model, and in our case, we lack sensor data for bathroom ventilation or humidity, making it challenging to capture relevant usage patterns. Therefore, integrating a physic-based reference in the loss function can help compensate for missing information. In contrast, our sensor features for bedrooms and living rooms capture most relevant effects, so adding the physics-based reference does not further enhance performance in these spaces.

\subsection{Ablation Study on Conformal Prediction}

\begin{figure}[h]
	\centering
	\includegraphics[width=1\linewidth]{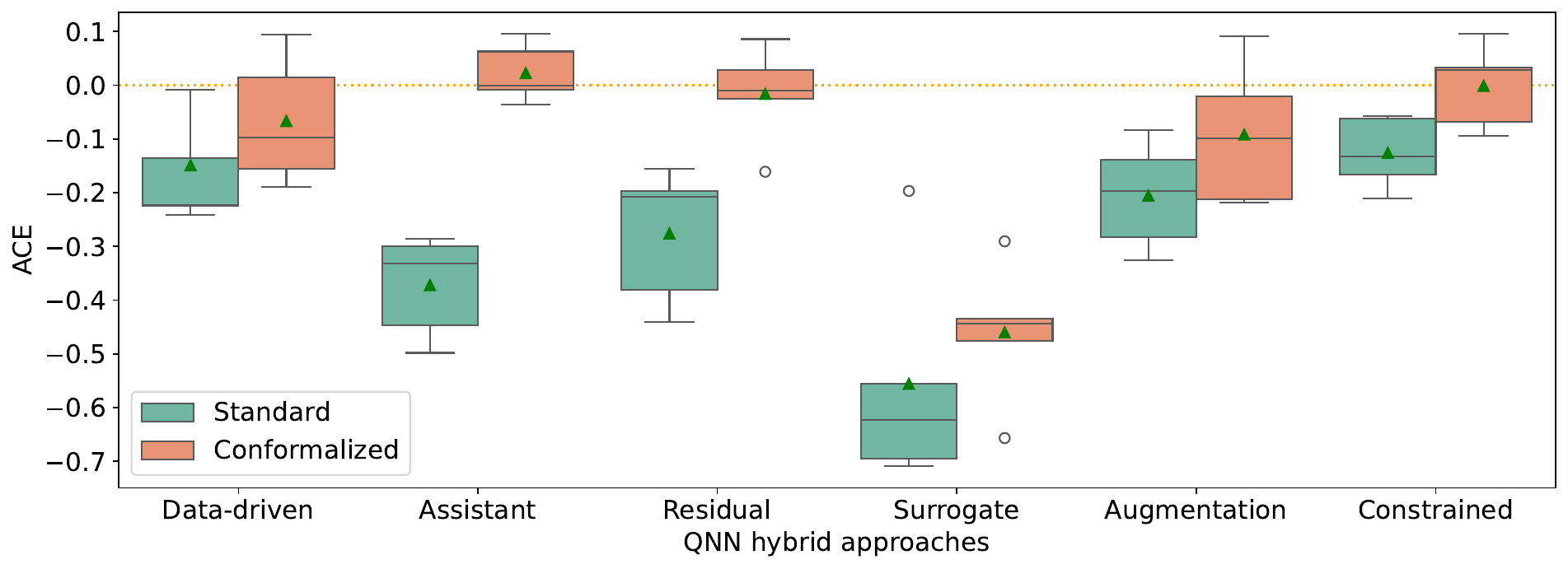}
	\caption{Comparison between ACE performance for the 90\% prediction interval of the QNN hybrid approaches with and without conformal prediction.}
	\label{fig:conformal_effect_ACE}
\end{figure}

\begin{figure}[h]
	\centering
	\includegraphics[width=1\linewidth]{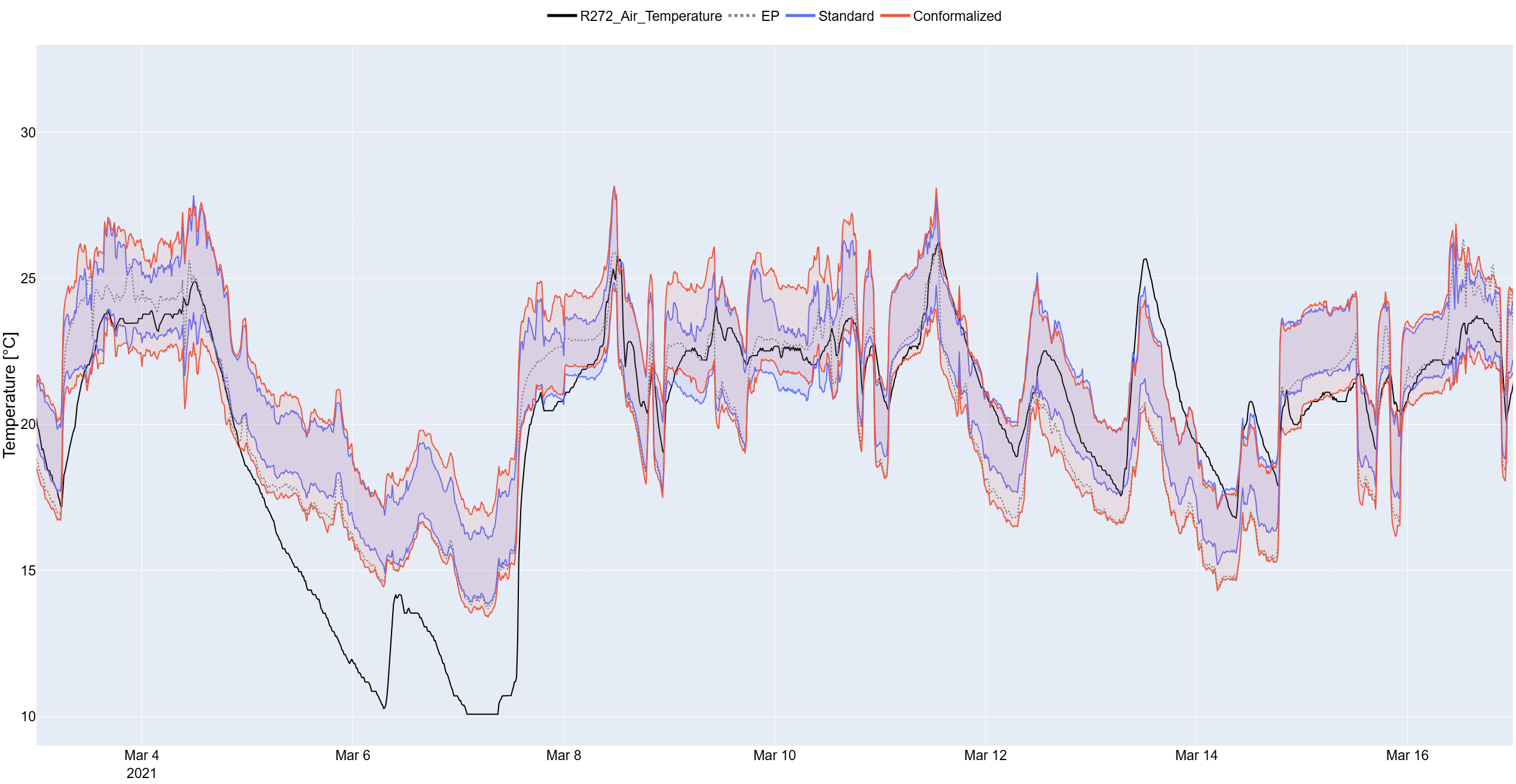}
	\caption{Forecast comparison between standard and Conformalized Quantile Regression for the Residual-QNN in the case of Bedroom 272.}
	\label{fig:conformal_effect_forecast}
\end{figure}

\begin{figure}[h]
	\centering
	\includegraphics[width=0.7\linewidth]{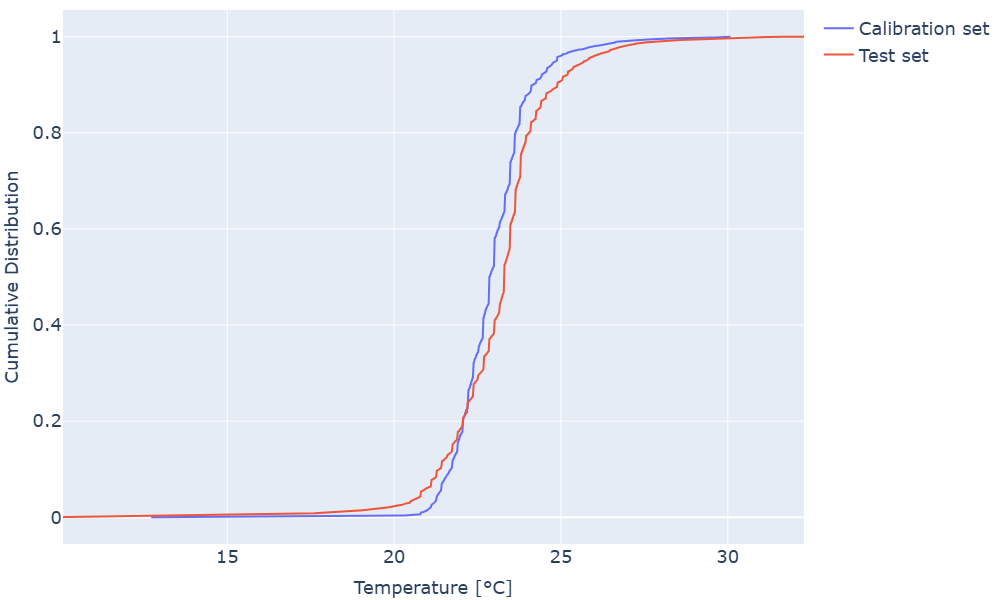}
	\caption{The Empirical Cumulative Distribution Function of the calibration and test set for the Temperature in Bedroom 272.}
	\label{fig:conformal_ecdf}
\end{figure}

As an additional ablation study, we investigate the impact of utilizing a conformal prediction framework compared to standard quantile regression. We focus our ablation on the 90\% prediction interval, given its widespread adoption in uncertainty quantification. Figure~\ref{fig:conformal_effect_ACE} presents a comparison of ACE values for the 90\% prediction interval in the case of the QNN. Our results indicate that conformal prediction improves ACE across both data-driven and hybrid approaches, attributable to its quantile calibration effect. The main effect of conformal prediction is a widening of the prediction intervals, as illustrated for the Residual-QNN in Figure~\ref{fig:conformal_effect_forecast}. This widening of the prediction interval width increases the probability of capturing the true value within the interval, thus reducing ACE. The effect is most pronounced in the Assistant approach, often leading to an overestimation of the interval width. Conversely, the data-driven model exhibits the smallest widening effect, with the interval width remaining underestimated even after correction.\\
It is important to note that we apply Conformalized Quantile Regression under the idealized exchangeability assumption -- that the calibration set and test set share the same distribution. However, in time series applications, distribution shift remains a significant challenge. In our case, factors such as seasonality and changing user behavior over time can partially violate the exchangeability assumption, as illustrated by the Empirical Cumulative Temperature Distribution Function for Bedroom 272 in Figure~\ref{fig:conformal_ecdf}. This issue could be mitigated by using a larger calibration dataset that spans all seasons, for example, by employing one full year of data only for quantile calibration.

\subsection{Prediction evaluation by quantile and prediction interval}

\begin{figure}[h]
	\centering
	\includegraphics[width=0.7\linewidth]{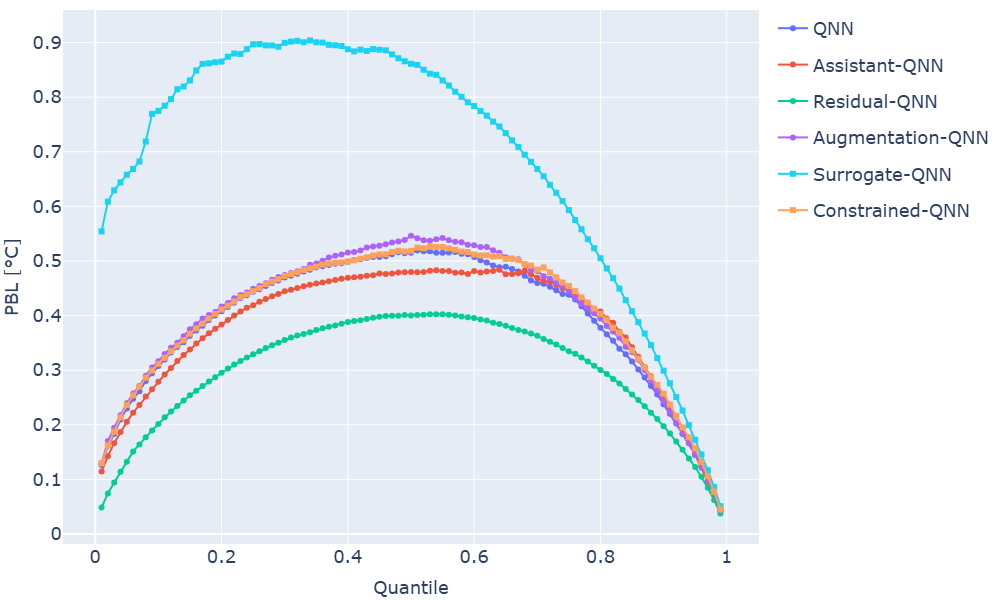}
	\caption{PBL per quantile for two QNN hybrid approaches in the case of Bedroom 272.}
	\label{fig:error_matrix_mean_R272_PBL}
\end{figure}

\begin{figure}[h]
	\centering
	\includegraphics[width=1.0\linewidth]{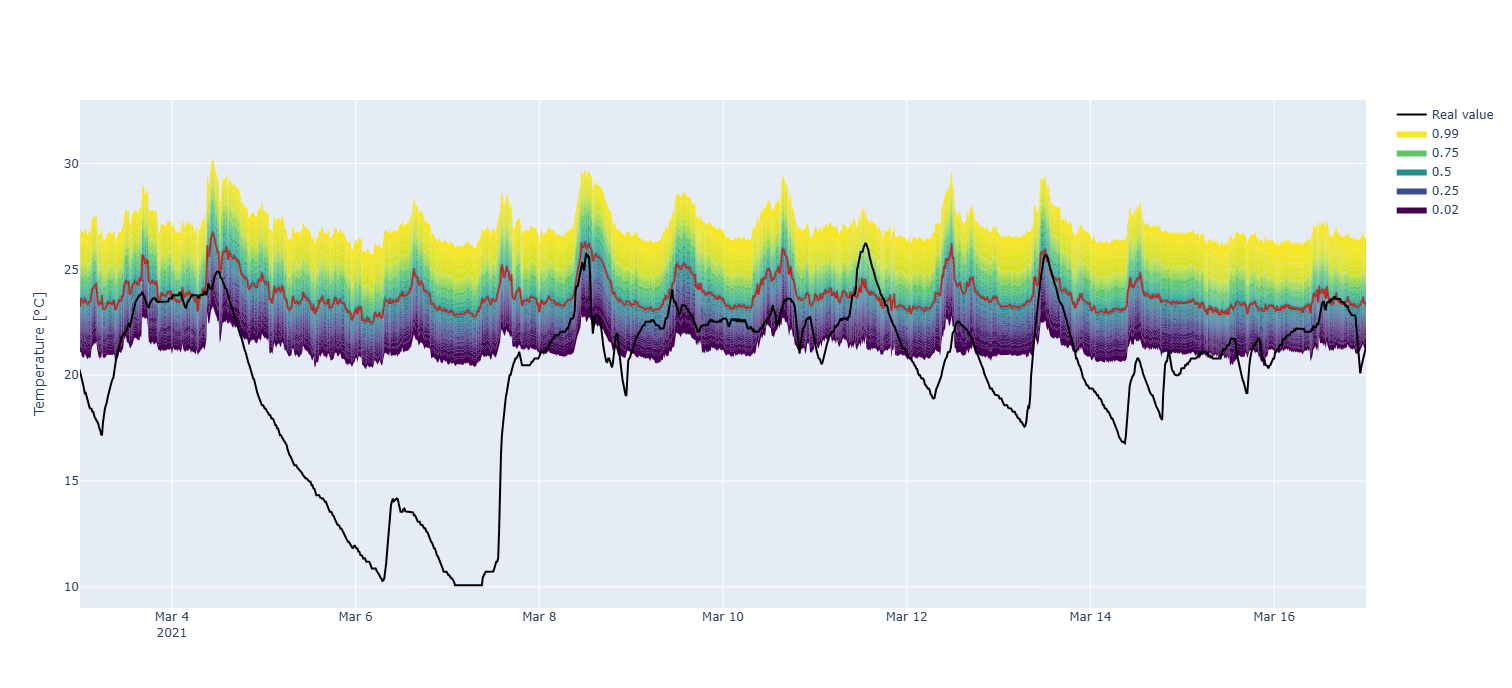}
	\caption{Purely data-driven QNN predictions of all 99 quantiles for selected time steps in March in the case of Bedroom 272. The median prediction is highlighted in red.}
	\label{fig:quantile_allcast_R272_MQFFNN}
\end{figure}

\begin{figure}[h]
	\centering
	\includegraphics[width=1.0\linewidth]{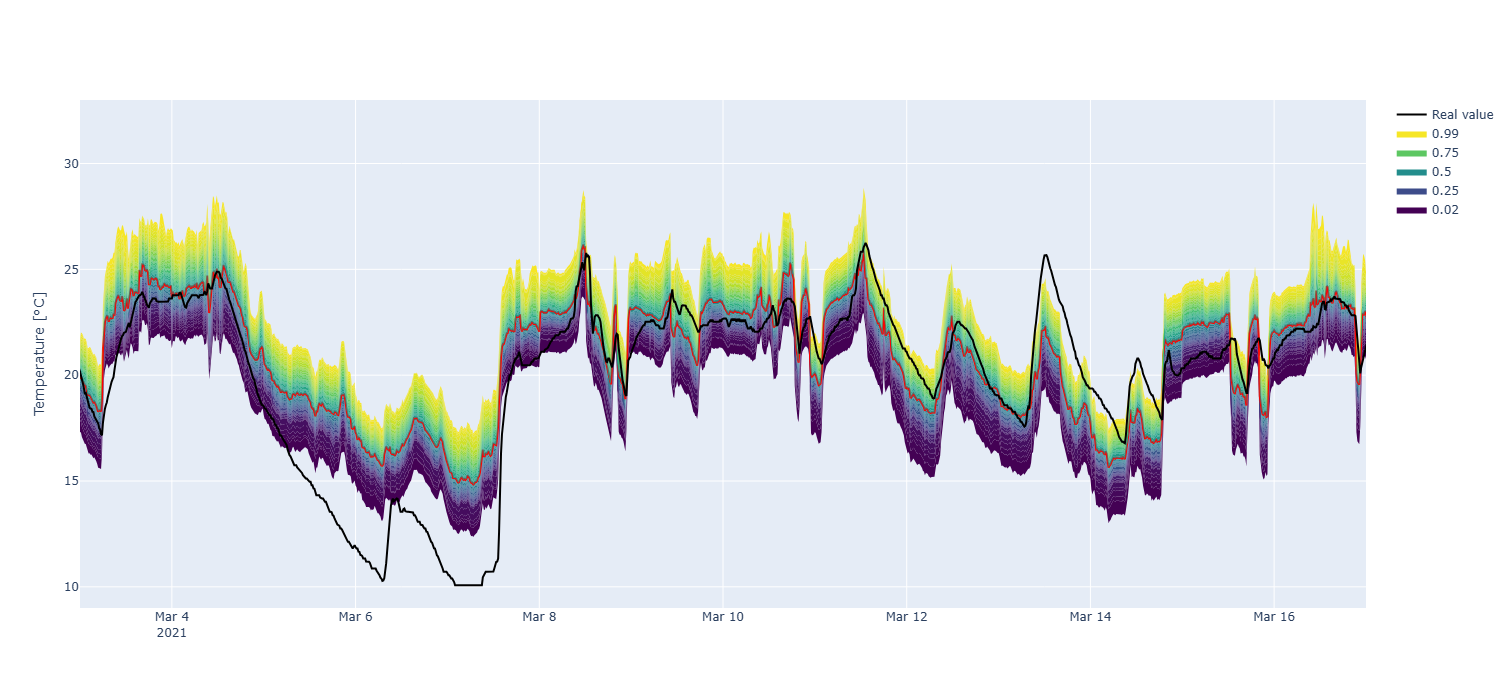}
	\caption{Residual-QNN predictions of all 99 quantiles for selected time steps in March in the case of Bedroom 272. The median prediction is highlighted in red.}
	\label{fig:quantile_allcast_R272_Residual-MQFFNN}
\end{figure}

The analysis in the preceding subsections were limited to either aggregated PBL values across all quantiles or focused specifically on the 90\% prediction interval for ACE and WKS. To enable a more comprehensive assessment of distributional performance, it is informative to plot error metrics across the full range of quantile and prediction intervals. Such visualization can yield valuable insights for practitioners interested in different levels of predictive uncertainty. Given the strong performance of the QNN hybrid approaches, our subsequent analysis will focus on these models. Additionally, we use Bedroom 272 as a representative case study, owing to its representative thermodynamic behavior. Figure~\ref{fig:error_matrix_mean_R272_PBL} presents the PBL for the QNN hybrid approach in Bedroom 272, broken down by quantile level. The Residual approach consistently achieves the lowest PBL across all quantiles, whereas the Surrogate approach yields the highest PBL. The data-driven, Assistant, Augmentation and Constrained methods produce intermediate and comparable results. The superior performance of the Residual-QNN is illustrated in Figure~\ref{fig:quantile_allcast_R272_MQFFNN} and \ref{fig:quantile_allcast_R272_Residual-MQFFNN}, which show that this model more closely tracks the observed temperature profile, particularly during sudden drops. By contrast, the purely data-driven QNN (Figure~\ref{fig:quantile_allcast_R272_MQFFNN}) exhibits less variation in its quantile predictions, resulting in higher PBL values across all quantiles.
The distinct inverse-quadratic shape observed in the PBL curves for most hybrid methods can be explained by the asymmetric nature of the pinball loss. Fore extreme quantiles, the loss function does not strongly penalize conservative predictions, as the true value often lies outside these ranges. However, for quantiles closer to the median, the pinball loss imposes more balanced penalties for both under- and over-prediction, and such situations occur more frequently. Another notable observation is that the Surrogate approach displays an asymmetric error profile, particularly at higher quantiles, indicating a systematic bias.

\begin{figure}[h]
	\centering
	\includegraphics[width=0.7\linewidth]{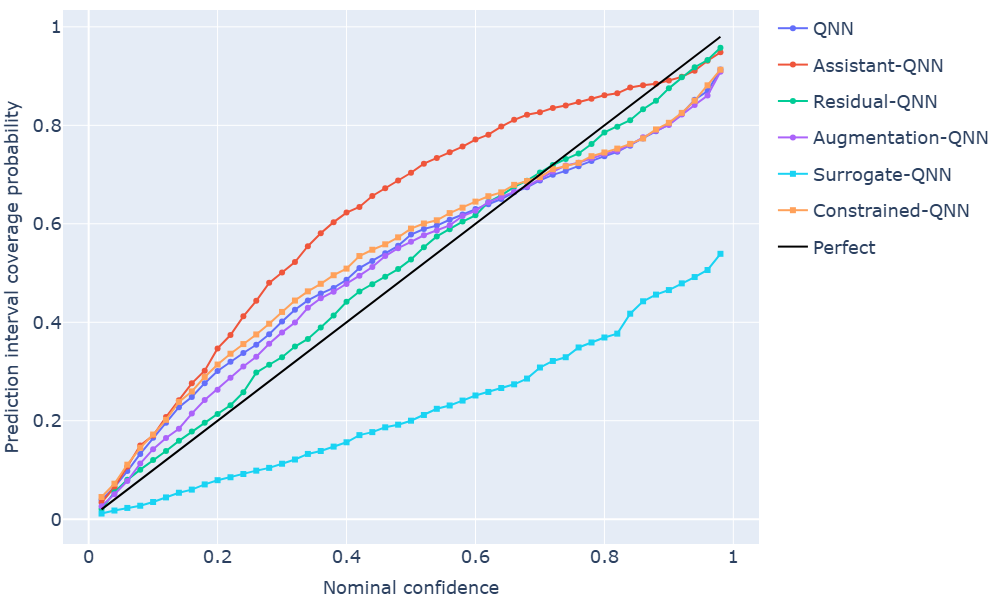}
	\caption{Reliability per prediction interval nominal confidence for two QNN hybrid approaches in the case of Bedroom 272.}
	\label{fig:error_matrix_mean_R272_REL}
\end{figure}

A similar pattern is observed in Figure~\ref{fig:error_matrix_mean_R272_REL}, which presents the reliability broken down by prediction interval for the QNN hybrid approaches in Bedroom 272. Reliability here is quantified using the components of the ACE metric, comparing the prediction interval coverage probability to the nominal confidence level, as defined in equation \ref{eq:ace}. The Residual approach stands out as the most reliable across all nominal confidences, exhibiting only slight overconfidence up to a nominal confidence of 0.7 and minimal underconfidence at higher levels. 
The purely data-driven, Augmentation and Constrained approaches display similar trends to the Residual model, but with larger deviations from perfect reliability. In contrast, the Assistant approach shows notable overconfidence across nearly all quantiles, while the Surrogate approach consistently demonstrates the lowest reliability, with substantial underconfidence across all nominal confidences. These findings are further supported by Figures~\ref{fig:quantile_allcast_R272_MQFFNN} and \ref{fig:quantile_allcast_R272_Residual-MQFFNN}, where the Residual approach is shown to generate quantile predictions that more closely follow the underlying thermodynamic patterns, resulting in better calibrated prediction intervals.
Another factor contributing to improved calibration is the asymmetric shape of the prediction intervals during sudden temperature drops, as seen around March 12th in Figure~\ref{fig:quantile_allcast_R272_Residual-MQFFNN}, where the predictive distribution becomes skewed towards the upper quantiles reflecting reduced uncertainty for higher temperatures. \\
While the Residual approach generally achieves the best calibration across the full predictive distribution, it is worth noting that during sudden temperature rises, such as those observed around around March 11th, the predictive distribution temporarily becomes overly narrow. Ideally, the quantile model should widen the spread towards the lower quantiles during sharp upward transitions to better capture the increased uncertainty associated with rising temperatures. These observations indicate that, although the Residual approach performs well overall, there remains room for improvement in capturing the dynamics of rapidly changing conditions.

\section{Conclusion}

In this work, we introduce probabilistic methods to five established hybrid approaches in building energy modeling combining physics-based and data-driven modeling alongside a purely data-driven method for thermodynamic quantile estimation. On this basis, we investigate the different probabilistic hybrid approaches extensively. Across multiple rooms, the Residual approach consistently outperforms the others in terms of average pinball loss, average coverage error, and Winkler score. In contrast, the Surrogate approach exhibits the weakest performance among the hybrid methods. Notably, the Residual model demonstrates a particular strength in capturing out-of-distribution behaviour, such as extended periods of window openings, a capability that was lacking in both the pure data-driven and other hybrid models. In addition, the Residual-QNN shows the best results for a representative Bedroom in terms of the pinball loss across all targeted quantiles and most reliable prediction intervals across all targeted prediction intervals. This can be partly explained by the asymmetric predictive distribution in case of sudden temperature drops, better representing the uncertainty of the downwards trend.
Nevertheless, even the Residual approach exhibits limitations during abrupt temperature increases with condensed predictive distribution, suggesting that further refinement is needed for highly dynamic scenarios. In case of the Constrained approach, we demonstrate that the increasing proportion of physics-based information is only beneficial for rooms that are harder to model due to unique dynamics and limited sensors coverage, such as bathroom.
This study highlights several promising directions for future research. The framework could be extended to address epistemic uncertainty, for example by incorporating Bayesian inference into the data-driven component. Another interesting avenue could be a sensitivity analysis of the physics-based submodel to provide further insight into epistemic uncertainty. Further, future work could include large-scale experiments encompassing a higher number of buildings, buildings with a higher number of rooms or extensions beyond the residential building type. Finally, the robustness of the conformal prediction procedure could be improved to better handle distribution shifts, potentially through adaptive conformal prediction methods.

\printcredits

\section*{Data availability}
The code and dataset will be made publicly available at the \href{https://github.com/Leo-VK/hybrid_bem}{GitHub repository}.

\section*{Acknowledgement}
We thank all involved members from the Urban Energy Systems Laboratory at Empa and Intelligent Maintenance and Operations Systems Laboratory at EPFL. Special thanks go to Fazel Khayatian for insightful discussions about Building Energy Modeling and the nestli platform. This project is funded by Empa research \& development grant 5213.00276.

\bibliographystyle{cas-model2-names}

\bibliography{cas-refs}

\clearpage

\end{document}